\documentclass{article}
\usepackage{spconf,amsmath,graphicx,xcolor}

\title{GPCGC: A Green Point Cloud Geometry Coding Method}
\name{Qingyang~Zhou \textsuperscript{1}, 
Shan~Liu \textsuperscript{2},
C.-C.~Jay~Kuo \textsuperscript{1}
\thanks{The authors acknowledge the gift support from the Tencent Media
Lab as well as the Center for Advanced Research Computing (CARC) at the
University of Southern California for providing computing resources that
have contributed to the research results reported within this
publication. URL: https://carc.usc.edu.}}

\address{University of Southern California, Los Angeles, California, USA$^1$\\
Tencent Media Lab, Palo Alto, California, USA$^2$}

\begin{document}
\ninept

\maketitle

\begin{abstract}

A low-complexity point cloud compression method called the Green Point
Cloud Geometry Codec (GPCGC), is proposed to encode the 3D spatial
coordinates of static point clouds efficiently. GPCGC consists of two
modules. In the first module, point coordinates of input point clouds
are hierarchically organized into an octree structure.  Points at each
leaf node are projected along one of three axes to yield image maps.  In
the second module, the occupancy map is clustered into 9 modes
while the depth map is coded by a low-complexity high-efficiency image
codec, called the green image codec (GIC). GIC is a multi-resolution
codec based on vector quantization (VQ). Its complexity is significantly
lower than HEVC-Intra. Furthermore, the rate-distortion optimization
(RDO) technique is used to select the optimal coding parameters. GPCGC
is a progressive codec, and it offers a coding performance competitive
with MPEG's V-PCC and G-PCC standards at significantly lower complexity. 

\end{abstract}


\section{Introduction}\label{sec:intro}

Point clouds have been widely used in computer-aided design, virtual
reality, autonomous driving, etc. Effective point cloud coding
techniques are critical to the storage and transmission of point cloud
data. This work examines the coding of the spatial coordinates of 3D
points of static point clouds, known as geometry compression, at lower
computational complexity while maintaining high coding performance. The
proposed solution is named the Green Point Cloud Geometry Codec (GPCGC). 

Point cloud geometry compression has been intensively studied in recent
years, including standardization and non-standardization activities.
V-PCC \cite{VPCC1} and G-PCC \cite{GPCC1} are two well-known point cloud
coding standards developed by the Moving Picture Experts Group (MPEG)
\cite{EmergingMPEGStandards,VPCCGPCCoverview}. V-PCC has the best coding performance for
dense point clouds among conventional (or non-learning-based) codecs.
Emerging learning-based codecs \cite{8803413,
AdaptiveDeepLearning-Based, wang2021TCSVT, borges2022fractional,
GeoCNNv2, SparsePCGC} exploit inter-sequence correlations and offer
impressive coding gains at the expense of higher computational
complexity. 

In this work, we propose a {\em low-complexity} learning-based codec for
point cloud geometry compression and call it the ``green point cloud
geometry codec" (GPCGC). It consists of three main ingredients: 1) a
hierarchical octree structure, 2) 3D-to-2D projection, which is similar
to V-PCC, and 3) coding of projected geometry maps via vector
quantization (VQ). GPCGC is a progressive and learning-based codec
because of the first and third ingredients, respectively.  Experiments
show that GPCGC offers a coding performance comparable with those of
MPEG's V-PCC at significantly lower complexity. 

\begin{figure*}[htb]
\begin{minipage}[b]{1.0\linewidth}
\centering
\centerline{\includegraphics[width=18.5cm]{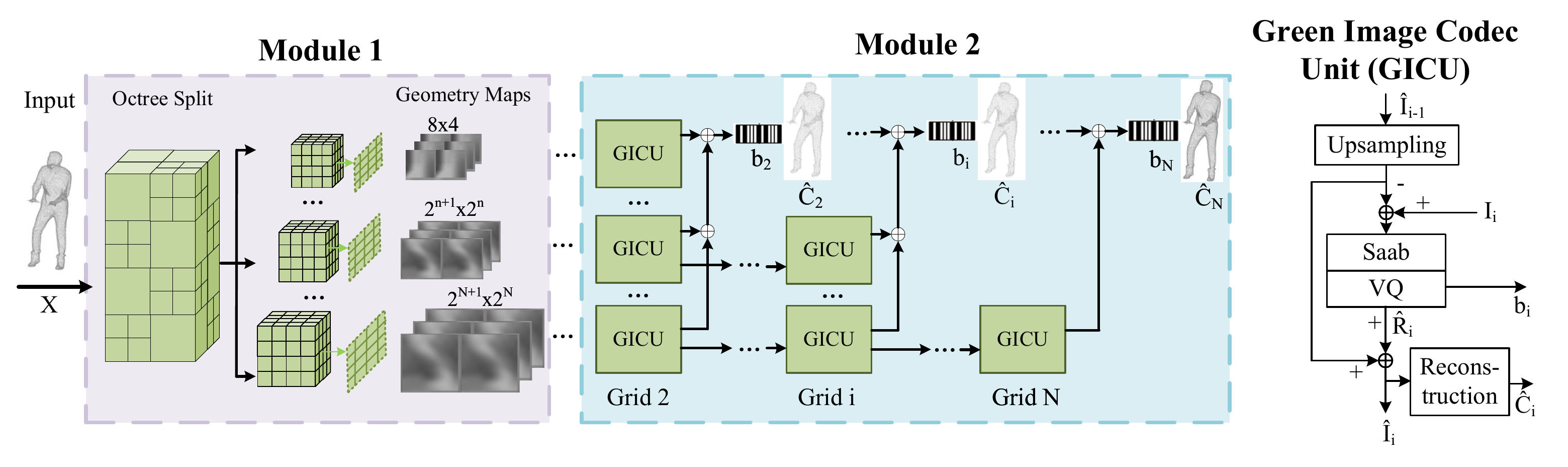}}
\caption{An overview of the proposed GPCGC method. The input point cloud
set is split hierarchically and projected into geometry maps in the first 
module. The geometry maps are coded by hierarchical GICUs in the second module. 
The GICU architecture is shown on the right.}\label{fig:framework}
\end{minipage}
\end{figure*}

\begin{figure}
\begin{minipage}[b]{1.0\linewidth}
  \centering
  \centerline{\includegraphics[width=8.5cm]{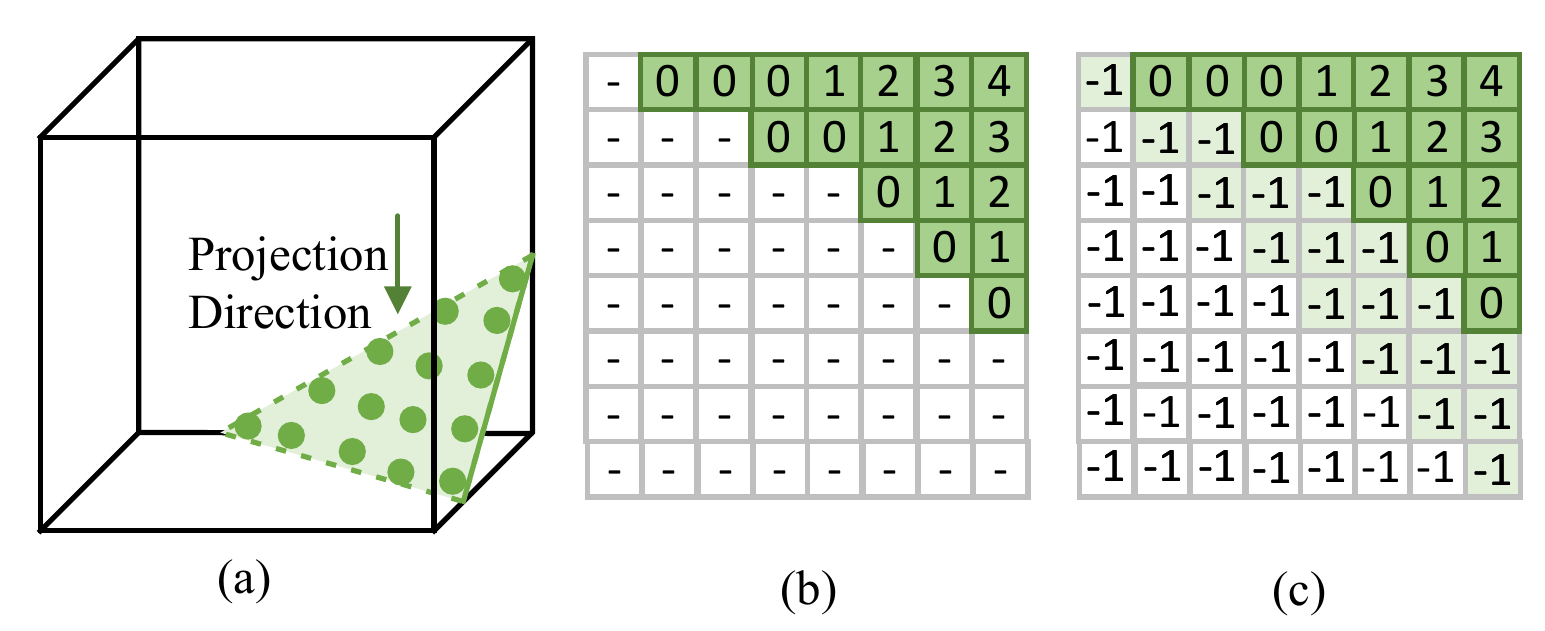}} 
\caption{Illustration of the depth map filling process: (a) the original 
3D points in a voxel, (b) the original depth map, (c) the modified depth
map, where the depth of empty pixels is labeled with ``-1". Only the
depths of green pixels are reconstructed based on the mode index
in depth decoding. Furthermore, the depth value of -1 is automatically 
filtered out.}\label{demo1}
\end{minipage}
\end{figure}

\section{Review of Previous Work}\label{sec:review}

V-PCC \cite{VPCC1} and G-PCC \cite{GPCC1} mean ``video-based point cloud
compression" and ``geometry-based point cloud compression",
respectively. They are both standards developed by MPEG.  V-PCC targets
the coding of dense point clouds. It uses a 3D-to-2D projection that
projects 3D points onto points in a 2D plane to yield a 2D
geometry/occupancy maps.  The latter is then encoded by the HEVC video
codec \cite{HEVC}. G-PCC aims at coding sparse point clouds and it
adopts an octree scheme to encode the 3D coordinates of points. 

Motivated by the success of deep-learning-based (DL-based) image coding
\cite{vae1,vae2}, researchers have applied the end-to-end optimized
Auto-Encoder (AE) to point cloud compression in \cite{8803413,
AdaptiveDeepLearning-Based, wang2021TCSVT}. They use 3D CNN or sparse
convolution-based AEs to represent the 3D occupancy model of voxelized
point cloud geometry data.  Inter-block and inter-sequence correlations
are extracted via block-based point cloud training.  DL-based methods
exhibit state-of-the-art coding performance at the expense of very high
computational costs and model sizes. The sparse convolution operator was
proposed in \cite{multiPCGC} to reduce the model size. Yet, several
inherent problems still exist (e.g., symmetric complexity of
encoders/decoders, one model for one bitrate.)

We attempt to leverage the advantages of V-PCC and learning-based codes in
the design of GPCGC. On one hand, GPCGC uses projection to reduce the
sparsity of points in the 3D space and map them into a 2D plane. On the
other hand, GPCGC encodes projected maps hierarchically with a
low-complexity learning-based method. For the latter, we revisit VQ. The
VQ technique can trace back to 80s \cite{gray1984vector}. Recent VQ
studies include: gain-shape VQ (called Daala)
\cite{valin2015perceptual,valin2016daala}, multi-grid multi-block-size
VQ (MGBVQ) \cite{wang2022lightweight}, and the green image codec (GIC)
\cite{wang2022green,greenlearningOverview}. Since VQ provides a low-complexity learning-based
solution, we adopt it in the proposed GPCGC method. 

\section{GPCGC Method}\label{sec:method}

The proposed GPCGC method consists of two main modules: 1) octree split
and projection and 2) depth map coding, as shown in Fig.
\ref{fig:framework}.  In the first module, a voxelized input point cloud
scan is split and projected under an octree structure. Its output is a
set of projected depth maps.  In the second module, depth maps are coded
by Green Image Codec Units (GICUs). These two modules are elaborated in
Sec. \ref{subsec:octree} and Sec. \ref{subsec:map_coding},
respectively. 

\subsection{Module 1: Octree Split and Projection}\label{subsec:octree}

{\em Octree Split.} Unlike V-PCC which decomposes a point cloud into
connected patches and generates global atlas maps, GPCGC adopts a
voxelization step that splits an input point cloud set into 3D
coarse-to-fine voxels hierarchically.  A parent voxel can be split into
8 child voxels by partitioning its side along the $x$, $y$, and $z$ axes
into two halves.  A recursive split operation can yield an octree
representation. An octree can be non-symmetric; namely, some branches can
go to a finer voxel level while others can stay at a coarser level. The
coarsest voxel number is $N_x \times N_y \times N_z$, where $N_x$,
$N_y$, and $N_z$ are user-selected parameters. The split level can be
determined by the rate-distortion optimization technique as discussed in
Sec.  \ref{subsec:RDO}. The projection from 3D spatial coordinates of a
point to its 2D coordinates is only conducted at each leaf node of the
octree.  This is the first major difference between our scheme and
V-PCC. 

{\em Projection.} By comparing orthographic projections along the $x$-,
$y$-, or $z$-axes, we choose the one that has the largest projection
area.  To measure the area, we discretize the projected 2D plane with a
uniform grid of $p \times p=p^2$ blocks. If a block does not contain any
projected point, its value is set to zero. Otherwise, its value is set
to one. The projection area is defined to be the total number of blocks
whose value is equal to one. The distances between 3D points in a leaf
voxel and the selected projection plane define a depth map.  A larger
project area is preferred since the corresponding depth map is flatter
and easier to encode. The depth map is also called a depth image.  The
projection direction at each leaf voxel is coded and included in the
bitstream. 

{\em Projection Ambiguity Resolution.} It is desirable that one block
has only one projected point. If this is the case, we can use the block
center location and the quantized depth to represent the corresponding
3D point. However, we may encounter two projection ambiguity problems.
First, we may see self-occlusions and hidden surfaces in a coarse voxel.
For this case, the voxel should be further split into 8 child voxels.
This process continues until such problems are resolved or the minimum
voxel size is reached. Second, two or more points are projected to the
same 2D blocks.  The ``thickness" concept is introduced to handle the
situation. That is, we generate two depth maps to store the maximum and
minimum distance values.  V-PCC treats the two maps as a two-frame video
to exploit of the inter-frame coding of HEVC.  We concatenate two maps
into a larger map for coding in GPCGC. 

\subsection{Module 2: Block Occupancy and Depth Coding}\label{subsec:map_coding}

{\em Block Occupancy Coding.} Some blocks in a geometry map may have no
projected points. V-PCC encodes the occupancy map losslessly. It is
expensive. To save the bit rate, we adopt a lossy coding scheme. It
classifies block occupancy patterns into full-occupied and 8
half-occupied cases (namely left/right, up/down, up-left/lower-right,
and up-right/lower-left), leading to 9 modes. We encode the optimal mode
that gives the best approximation to the underlying pattern at each leaf
node and include it in the bit stream.  One example is given in Fig.
\ref{demo1}, where we choose the upper-right triangle mode and fill
empty blocks with dummy depth value ``-1" to simplify the depth
encoding/decoding procedure.  After depth decoding, only the depths of
green pixels are reconstructed based on the mode index and the
dummy depth value is automatically filtered out.  When the range of
depth value lies in $[0, D_{\max}]$, the dummy depth value is set to
$-1$ or $D_{\max}+1$ so as to maintain a smooth depth map defined on a
squared region. 

The lossy coding of block occupancy allows a very small bit rate but
could introduce distortion.  This is the second major difference between
our scheme and V-PCC. The depth value of an empty block that is
surrounded by non-empty blocks partially or fully can be inferred from
the value of its nearest non-empty block (or the averaged value if there
are multiple ones). This is equivalent to adding a new point to the
point cloud set.  The occupancy distortion can be mitigated by such
a process. 

{\em Depth Coding.} Since a depth map is fundamentally a gray-scale
image, it can be coded by any image codec such as JPEG, H.264-Intra,
HEVC-Intra. We adopt a low-complexity learning-based image codec known
as the GIC method \cite{wang2022green}. Its complexity is significantly
lower than HEVC-Intra, which meets our low complexity requirement.
However, its coding gain is worse than that of HEVC-Intra. The poorer
coding gain can be compensated by the effectiveness of block occupancy
coding as described above. It is shown in Sec. \ref{sec:experiments}
that the RD performance of our GPCGC is comparable with V-PCC.

GIC downsamples an input image into several spatial resolutions from
fine-to-coarse grids and computes image residuals between two adjacent
grids. Then, it encodes the coarsest content, interpolates content from
coarse-to-fine grids, encodes residuals, and adds residuals to
interpolated images for reconstruction.  All coding steps are
implemented by VQ while all interpolation steps are conducted by the
Lanczos interpolation.  To facilitate VQ codebook training, a
data-driven transform, called the Saab transform
\cite{kuo2019interpretable,cwsaab} is applied for energy compaction and, thus,
dimension reduction. We can express the whole GIC framework as the
cascade of multiple GIC units, where each GIC unit (GICU) is applied at
a fixed grid level. The structure of each GICU is shown in the right of
Fig. \ref{fig:framework}.  This is the third major difference between
our scheme and V-PCC. 

\subsection{Rate Control via RDO}\label{subsec:RDO}

We revisit the octree decomposition problem in Module 1. Suppose no
projection problem is encountered. A larger voxel size demands a lower
coding bit rate but has a higher distortion.  The rate-distortion
optimization (RDO) technique is used to determine the optimal split. 
The RDO cost function can be expressed as
\begin{equation}\label{lagrangian}
C_n = D_n + \lambda_n R_n, 
\end{equation}
where $C_n$, $D_n$, $R_n$, and $\lambda_n$ denote the cost, the bit rate,
the distortion, and the Lagrangian multiplier at the $n$th split,
respectively. 

{\em Rate Modeling.} The bit rate is mainly determined by VQ's coding
indices.  Let $R_{n}$ denote the total number of bits to encode the
depth map at the $n$th split level. Then, we have
\begin{equation}\label{eq:RD}
R_{n}= \sum_{i} \log_2 {C_{n,i}},
\end{equation}
where $C_{n,i}$ is the
codebook size of the $i$th grid GICU at the $n$th split level.

{\em Distortion Modeling.} V-PCC uses depth coding distortion to
evaluate the local distortion. Here, we use the point-to-point 
Hausdorff distance to measure the local distortion:
\begin{equation}\label{eq:D}
D_n= \max(\cfrac{1}{N_{P1}}\sum_{i} \lVert E(i,j)\rVert^{2}, 
\cfrac{1}{N_{P2}}\sum_{j} \lVert E(j,i)\rVert^{2}),
\end{equation}
where $P1$ and $P2$ denote the input and reconstructed point cloud sets
at the $n$ split, $i$ and $j$ are points of $P1$ and $P2$, respectively,
and $\lVert E(i,j)\rVert^{2}$ is the Hausdorff distance between points
$i$ and $j$. 

To allow various trade-offs between $R_n$ and $D_n$, we can adjust
$\lambda_n$ flexibly. The value of $\lambda_n$ is larger for a smaller
$n$.  This is different from end-to-end DL-based coding methods that
have a fixed $\lambda$ in the loss function. We compare the two cost
functions, $C_n$ and $C_{n+1}$, and choose the optimal one.  The RDO
process is the fourth major difference between our scheme and V-PCC. 

\begin{figure*}[htb]
\begin{minipage}[b]{1.0\linewidth}
  \centering
  \centerline{\includegraphics[height=5.8cm, width=18.0cm]{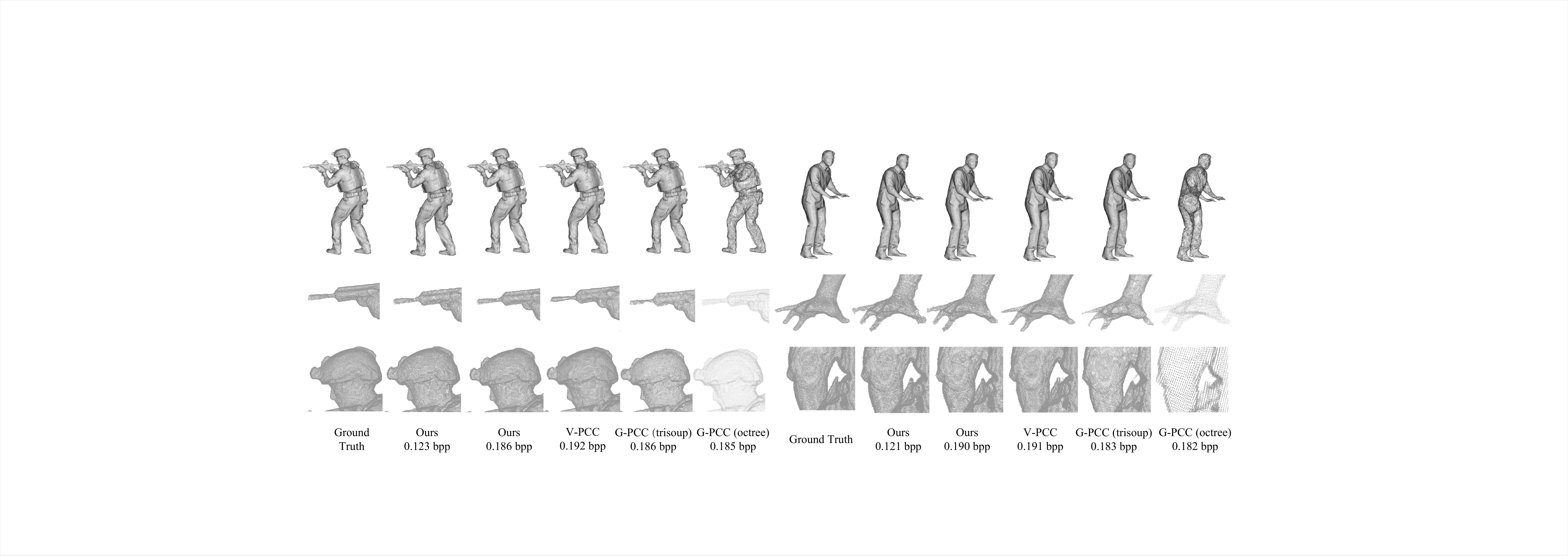}}
\caption{Visual Comparison of \textit{soldier\_vox10\_0690} and
\textit{loot\_vox10\_1200}: the whole PC view and two zoom-in views
(from top to bottom) of ground truth, GPCGC at two bitrates, V-PCC,
G-PCC (trisoup), and G-PCC (octree) (from left to
right).}\label{subjective}
  \centering
  \centerline{\includegraphics[width=18.5cm]{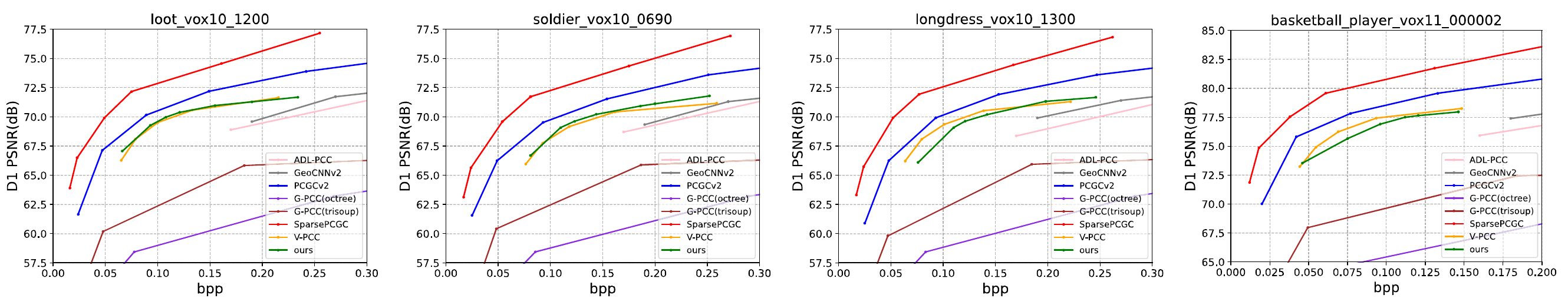}}
  \centerline{\includegraphics[width=18.5cm]{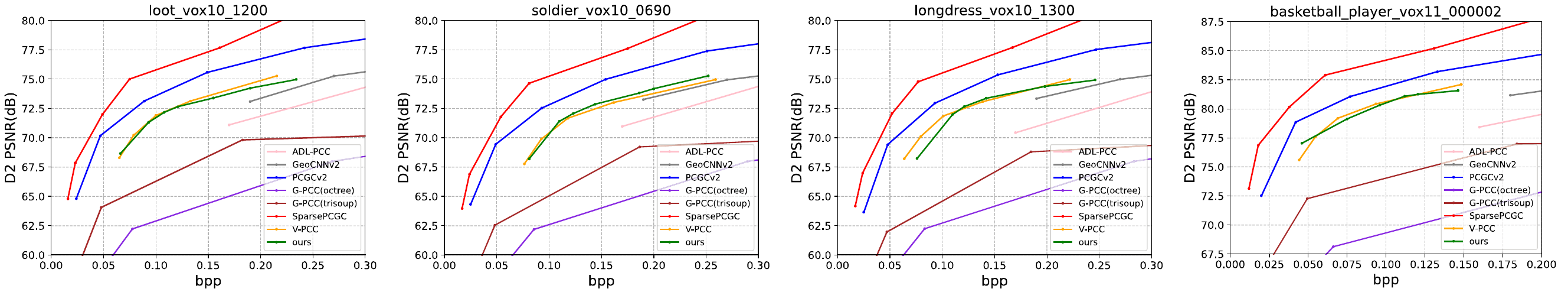}}
  \label{rdcurve}
\caption{RD Performance comparison of eight coding methods: V-PCC, G-PCC
(octree), G-PCC (trisoup), ADL-PCC, GeoCNNv2, PCGCv2, SparsePCGC, and
GPCGC (ours).}\label{fig:RD2}
\end{minipage}
\end{figure*}

\section{Experiments}\label{sec:experiments}

{\em Experimental Setup.} The coarsest input voxel size is 32x32x32.
The split number ranges from $n=0$ to $3$. When $n=3$, the voxel size is
4x4x4. The Lagrangian multipliers, $\lambda_n$, in Eq. (\ref{eq:RD}) are
empirically set to $\lambda_3=\lambda$, $\lambda_2=2.5 \lambda$,
$\lambda_1=4.6 \lambda$, $\lambda_0=8.0 \lambda$. The coding bit rate is
controlled by $\lambda$. We use mesh models from ShapeNet
\cite{shapenet} and SHREC'19 \cite{SHREC2019} as the training data.
They are converted to point clouds by uniform sampling and voxelized on
a 512x512x512 occupancy space. Then, point clouds are split into cubes
of various sizes (e.g., 32x32x32, 16x16x16, 8x8x8, and 4x4x4). 20,000
cubes are selected to train our model, and each GICU is trained by 2000
samples on average. The training process is implemented from coarse to
fine grids. The VQ module in the GICU is trained using the faiss.KMeans
method \cite{faiss}. The testing set includes single frames from the 8iVFB
\cite{8ipeople} and the Owlii \cite{owlii} point cloud datasets. The
experiment follows the MPEG Common Test Condition \cite{CTCCondition}. 

{\em Performance Analysis.} We compare our codec with seven other
codecs: V-PCCv18.0 with HM video encoder\cite{website:VPCC} , G-PCCv14.0 (trisoup)\cite{website:GPCC}, G-PCCv14.0 (octree)\cite{website:GPCC}, ADL-PCC
\cite{AdaptiveDeepLearning-Based}, GeoCNNv2 \cite{GeoCNNv2}, PCGCv2
\cite{multiPCGC}, and SparsePCGC \cite{SparsePCGC}.  The last four are
DL-based methods. The bit rate is calculated using bits per input point
(bpp).  We use the BD-rate to evaluate the coding performance.  The
distortion metrics are the mean-squared error (MSE) of the
point-to-point (p2point or D1) and the point-to-plane (p2plane or D2)
distances. Their RD curves for four test sequences are shown in Fig.
\ref{fig:RD2}. We categorize the eight methods based on their RD
performance from the best to the worst RD into four groups: 1)
SparsePCGC and PCGCv2, 2) GPCGC (ours) and V-PCC, 3) GeoCNNv2 and
ADL-PCC, and 4) G-PCC (trisoup) and G-PCC (octree).  Although there is a
performance gap between our codec and two state-of-the-art DL methods,
SparsePCGC and PCGCv2, the model size and complexity of our codec are
significantly lower than those of the two as discussed below.  Two coded
point cloud sets and their zoom-in views of GPCGC, V-PCC, G-PCC
(trisoup), and G-PCC (octree) are shown in Fig. \ref{subjective} for
visual comparison.  Our method offers a smooth transition from a higher
bit rate (i.e., 0.18bpp) to a lower bit rate (i.e., 0.12bpp) in terms of
subjective quality. 

{\em Model Size and Complexity Analysis.} We compare the model size and
the encoding/decoding floating-point operations (FLOPs) of the top four
performers (i.e., SparsePCGC, PCGCv2, V-PCC, and our GPCGC) in Table
\ref{tab:FLOPs}. V-PCC is not a learning-based codec. Its model size is
negligible. Our method can handle multiple coding rates with a single
model.  It has a size of 0.33 million parameters. In contrast, DL-based
codecs need multiple models to handle multiple bit rates. The table only
lists the size of one model. If there are $K$ models in the model zoo,
the actual model sizes are $0.78 \times K$ and $2.88 \times K$ million
parameters for PCGCv2 and SparsePCGC, respectively. 

The encoding/decoding FLOPs are computed under the following setting.
The input point clouds are voxelized into 1024x1024x1024 voxels. The
occupancy ratios of the vox10 test sequences range from 0.071\% to
0.101\% with an average value of 0.084\%.  For V-PCC, we use the Intel
VTune software to measure their FLOPs in the geometry coding process
over the same sequences with 10-bit precision. For DL-based methods, we make an
estimation based on their network structure.  We report the averaged
encoding/decoding FLOPs in the last two columns of Table
\ref{tab:FLOPs}.  As shown in the table, our codec has 4G and 64M FLOPs
on average for the encoding and decoding of a point cloud set,
respectively. In comparison, DL-based methods have significantly higher
encoding/decoding complexity. Their decoding complexities are about 1000x of
ours. Moreover, the encoding/decoding FLOPs of our codec are smaller
than those of V-PCC. The savings are 60\% and 64\%, respectively.  We
should point out that FLOPs only account for a very small percentage of
micro-Operations in traditional codecs such as V-PCC and G-PCC. If we
take the non-floating-point operations into account, our codec has 
even more significant savings in complexity.  The low FLOPs of our
decoder come from the simple structure of GIC \cite{wang2022green}.
Only look-up tables are required for VQ decoding and some matrix
multiplications are needed for the inverse-Saab transform. 

\begin{table}[t]
\centering
\begin{tabular}{c c c c} \hline
Methods  & Model Size & Enc FLOPs & Dec FLOPs  \\ \hline
PCGCv2 & 0.78M & 17.5G & 60G\\
SparsePCGC  & 2.88M & 35G & 100G \\
V-PCC & - & 10G & 180M\\
GPCGC (ours) &  \textbf{0.33M} & \textbf{4G} &\textbf{64M}\\ \hline
\end{tabular}
\caption{Comparison of the model sizes and the complexity in terms of
the encoding/decoding floating-point operations (FLOPs) of four benchmarking
codecs.} \label{tab:FLOPs}
\end{table}

\section{Conclusion}\label{sec:conclusion} 

A low-complexity point cloud geometry compression method, called the
Green Point Cloud Geometry Codec (GPCGC), was proposed.  The novel
contributions include projection on octree-decomposed voxels, lossy
occupancy map coding, GICU-based depth map coding, and RDO for bit rate
control. The last one is difficult to achieve in DL-based codecs.  As
compared with MPEG V-PCC, GPCGC achieves lower encoding/decoding
complexity (with a saving of around 60\%) while maintaining a comparable
coding gain. As compared with DL-based models, GPCGC has significantly
smaller model parameters and lower encoding/decoding complexity.  The
low-complexity advantage of GPCGC comes from the multi-resolution depth
map coding with GICU, which consists of the Saab transform and VQ. We
plan to develop a green coding solution for dynamic point clouds as an
extension in the future. 

\newpage
\bibliographystyle{IEEEtran}
\bibliography{refs}

\end{document}